# Linear Growth of Spiral SASI Modes in Core-Collapse Supernovae

John M. Blondin and Samantha Shaw
*Department of Physics, North Carolina State University, Raleigh, NC 27695-8202*

## ABSTRACT

Two-dimensional axisymmetric simulations have shown that the post-bounce accretion shock in core collapse supernovae is subject to the Spherical Accretion Shock Instability, or SASI. Recent three-dimensional simulations have revealed the existence of a non-axisymmetric mode of the SASI as well, where the postshock flow displays a spiral pattern. Here we investigate the growth of these spiral modes using two-dimensional simulations of the post-bounce accretion flow in the equatorial plane of a core-collapse supernova. By perturbing a steady-state model we are able to excite both one, two and three-armed spiral modes that grow exponentially with time, demonstrating that these are linearly unstable modes closely related to the original axisymmetric sloshing modes. By tracking the distribution of angular momentum, we show that these modes are able to efficiently separate the angular momentum of the accretion flow (which maintains a net angular momentum of zero), leading to a significant spin-up of the underlying accreting proto-neutron star.

## 1. Introduction

The post-bounce phase of core-collapse supernovae has drawn increased scrutiny in recent years (see Mezzacappa (2005) for a review), with a strong emphasis on the role of multidimensional effects such as rotation and convection. Following the gravitational collapse and subsequent bounce of the inner core of a massive star, the nascent supernova shock stalls out at a radius of order 100 km. Over the next few hundred milliseconds, the shock is revived and the supernova initiated by an as yet undetermined mechanism. Neutrino heating of the post-shock flow by neutrinos emanating from the proto-neutron star is believed to be important to the subsequent explosion, but other effects including convection and the instability of the accretion shock itself may play a critical role.

The Spherical Accretion Shock Instability (SASI) was discovered in two-dimensional axisymmetric simulations by Blondin, Mezzacappa, and DeMarino (2003, hereafter Paper I). These simulations were constructed to understand the origins of the turbulent fluid dynamics in the neutrino heating, or ``gain,'' region between the proto-neutron star surface, or ``neutrino-sphere,'' and the shock during the critical neutrino-heating epoch after stellar core bounce. Despite the fact that their initial conditions were stable to convection, the post-shock flow quickly became turbulent, driven instead by an instability of the stalled accretion shock. They found that the SASI is dominated by a sloshing mode characterized by a spherical harmonic, $Y_{l,m}(\theta,\phi)$, of $l=1$, and demonstrated that the SASI could have a dramatic impact on both the generation and the morphology of the supernova.

A quantitative study of the SASI in the linear regime by Blondin and Mezzacappa (2006, hereafter Paper II) suggests that the SASI is a standing acoustic wave in the post-shock cavity between the proto-neutron star surface and the shock, and that this standing wave is fed by the dynamic response of the accretion shock. In this scenario, the high pressure of the acoustic wave pushes the accretion shock outward by some small distance. Although the ram pressure at this new displaced shock location is lower than at the equilibrium shock radius (varying as $r^{-2.5}$), the steep post-shock pressure profile of $r^{-4}$ means that the pressure at a fixed radius will increase if the accretion shock is displaced outwards. This provides a positive feedback, leading to an exponential growth of the global pressure wave.

Alternatively, the SASI may be the result of a vortical-acoustic instability (Foglizzo et al. 2006), as originally suggested in Paper I. In this scenario velocity perturbations generated at the accretion shock are advected radially inward with the accretion flow. At some radius deep below the accretion shock these perturbations generate pressure waves that propagate radially outward and push back on the shock, starting the cycle over again.

The presence of the sloshing mode has been corroborated by several independent numerical models. Ohnishi et al. (2005) presented 2D simulations including the effects of neutrino heating (and hence post-shock convection), and found SASI growth curves virtually identical to those of Paper I. Perhaps the most striking evidence that the SASI can influence a supernova is given by Janka et al. (2004), who presented an 11 $M_\odot$ model computed on an axisymmetric grid spanning π radians. This simulation produced a mild explosion with a clear *l=1* asymmetry, while a comparable model computed on a grid spanning only *π/2* radians did not explode (Buras et al. 2003). Recent linear perturbation analyses (Foglizzo et al. 2006; Laming 2006) provide additional evidence for the SASI.

The existence of the SASI in three dimensions (without the artificial restriction of axisymmetry) was confirmed by a series of simulations by Blondin (2005). In addition to showing that the $l=1$ mode was still the dominant mode in 3D, these simulations found that the transition to the nonlinear phase of the SASI was typically marked by a loss of axisymmetry and the formation of a very robust spiral wave mode (*m=1*). Furthermore, this non-axisymmetric spiral wave led to the accretion of significant angular momentum, perhaps sufficient to explain the rapid rotation of young radio pulsars left behind by core collapse supernovae (Blondin and Mezzacappa 2006).

In this paper we seek a better understanding of the non-axisymmetric modes of the SASI. Are these new modes linearly unstable? How do they compare to the axisymmetric modes studied in Paper II? How does the nonlinear phase of these spiral modes lead to the accretion of angular momentum? We address these questions by describing a 2D time-dependent numerical model that can be used to evolve non-axisymmetric modes in section 2. The linear growth of these modes is quantified in section 3, and the resulting rotational flow in the nonlinear epoch is described in section 4.

## 2. Non-Axisymmetric Model of a Standing Accretion Shock

To study non-axisymmetric modes of the SASI we employ the same numerical model as in Paper II, but here we use a 2D polar grid spanning the entire equatorial plane ($\phi$ varies from 0 to $2\pi$) rather than a spherical-polar grid stretching from one pole to the other ($\theta$ varies from 0 to $\pi$). The grid volume still varies as $r^3$ to allow the proper steady-state solution (i.e., the 2D computational grid is a wedge in the equatorial plane rather than a flat disk as in a cylindrical-polar grid). This grid geometry does not correspond to any symmetry in three dimensions, but instead is an approximate way of modeling what is necessarily a fully 3D flow.

The initial model is described by a steady-state spherical accretion shock at a radius $R_s$, and an inner reflecting boundary at the surface of the proto-neutron star, $r_*$. The steady nature of this model is maintained by a constant mass accretion rate onto the shock and a thin cooling layer near the surface of the PNS removing the accretion energy. The cooling parameters used in this paper are $\alpha = 3/2$ and $\beta = 5/2$, and the cooling is turned off when the temperature reaches a floor value of 1/6 of the post-shock temperature. At this value the scale height of an isothermal atmosphere is comparable to the zone spacing of the grid. Over time the density builds up in the first radial zone outside the surface of the PNS. When the gas in this numerical zone cools to the point where its entropy drops below a value of 1% of the post-shock entropy, we assume that gas is accreted onto the PNS and we remove enough mass (and associated angular momentum) from the computational grid to raise the entropy back up to the floor value (keeping pressure constant). By tracking this material we can measure the mass and angular momentum accreted onto the PNS.

Given that the SASI is an instability of the accretion shock (Paper II), the linear evolution of this model should be relatively insensitive to the details of the boundary layer at the surface of the accreting PNS. To test this assertion we have run several models with different inner boundary conditions, including dropping the value of the entropy at which dense gas is taken off the grid, decreasing the floor temperature of the cooling function, and changing the boundary conditions on the transverse velocity. The models shown here use the traditional reflecting boundary condition for an inviscid flow, namely setting the gradient of the velocity component parallel to the surface to zero. We also ran simulations where the parallel component of the velocity was set to zero. In all of these cases, the growth of the SASI through the linear regime was unchanged. Even if no mass was removed from the grid, the only significant difference was the build-up of a very dense, isothermal (at the cooling cut off temperature) atmosphere in the first few radial zones of the grid. In spherical symmetry the accretion shock would slowly creep outward, but this was sufficiently slow that it had only a minor effect on the linear growth of the SASI.

The simulations presented here were run on a grid of $450^2$ zones. Both higher and lower resolution runs were performed to confirm the convergence of these solutions. Further details of this model can be found in Paper II. We use the results of Liebendoerfer et al. (2001) to scale our models to relevant supernova values: $R_s$ = 230 km, 1.2 M_\odot interior to the shock, and a mass accretion rate of 0.36 M_\odot s$^{-1}$.

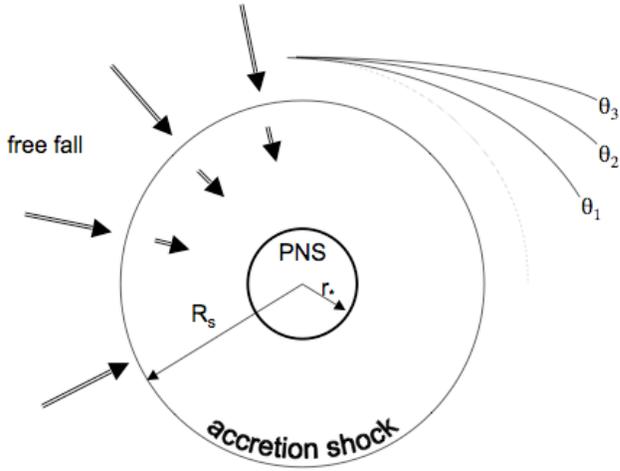

**Figure 1.** Schematic diagram of the spherical accretion shock model and the density perturbations used to excite non-axisymmetric modes of the SASI. The orientation of the bar, θ, is varied to empirically find the best value for exciting the desired spiral mode.

We can excite axisymmetric modes with this model by adding a density perturbation of the form $\cos(l\phi)$ to the free-falling gas above the accretion shock. The resulting growth rates are in good agreement with those measured in Paper II using an axisymmetric spherical-polar grid, although the periods measured here are ~ 15% larger than those measured in Paper II. This difference is likely the result of the slightly different flow geometry. As a wave propagates from one side of the shock cavity to the other, it will experience more of a divergence as it approaches the equator and convergence as it approaches a pole on the spherical-polar grid compared to the equatorial grid used in this paper.

To excite non-axisymmetric modes we drop an overdense bar (typically less than 1% denser than the surrounding accretion flow) tilted with respect to a surface of constant radius by an angle θ. This geometry is illustrated in Figure 1. We ran several simulations with various values of this drop angle in an effort to find the cleanest excitation of a single spiral mode. By dropping multiple bars we are also able to excite multi-armed spirals ($m=2,3$).

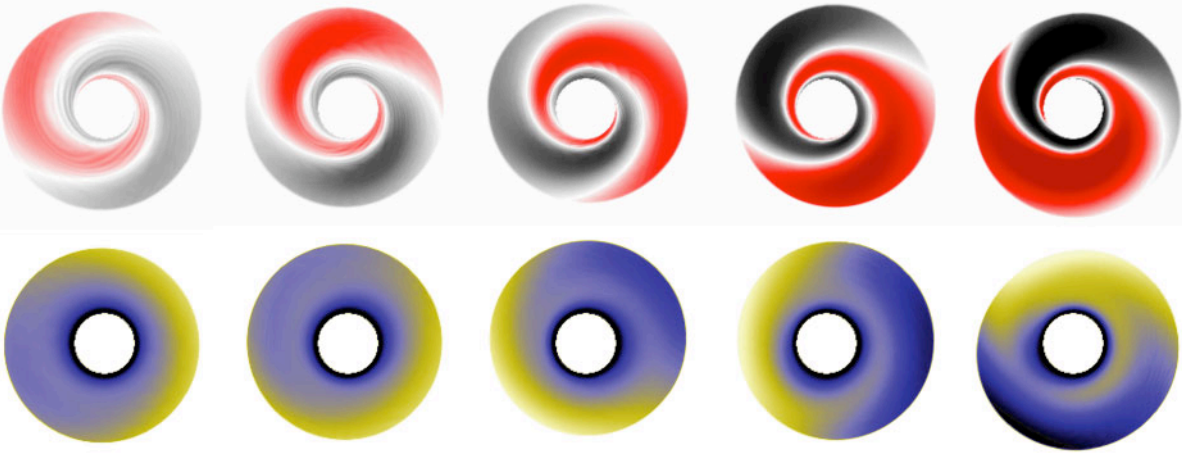

**Figure 2.** The evolution of the $m=1$ spiral mode of the SASI is illustrated here in a sequence of images from a simulation with $r_* = 0.3\ R_s$. The top row shows the angular component of the velocity, with black corresponding to clockwise rotation and red to counter-clockwise. The bottom row shows the gas pressure, with blue (white) representing values higher (lower) than equilibrium. We have removed the strong radial dependence of the equilibrium pressure profile by plotting $Pr^4$.

## 3. Linear Growth of Spiral Modes

An example of a spiral SASI mode is illustrated in Figure 2, where we show both the angular component of the velocity and the pressure over one period of rotation of the spiral wave. Throughout this image sequence the accretion shock remains nearly spherical, indicative of the linear phase of the SASI. The spiral pattern seen in the velocity is created by the radial advection of the velocity perturbation imprinted on the flow by the perturbed spherical shock. As the shock perturbation, which can be thought of as a wave on the shock, travels in a clockwise direction along the surface of the shock it generates a nearly symmetric velocity perturbation on the immediate post-shock flow as a result of the slight shock obliquity induced by the traveling wave. This velocity perturbation drifts inward with the accretion flow, shearing the perturbation into a spiral pattern.

The pressure images show a high pressure in one hemisphere and a low pressure in the other, with the maximum deviation occurring near the accretion shock. The velocity and pressure patterns rotate synchronously around the accreting star with a rotation period of 60 msec. This period matches the oscillation period of the $l=1$ sloshing mode of the SASI as measured with this same numerical model.

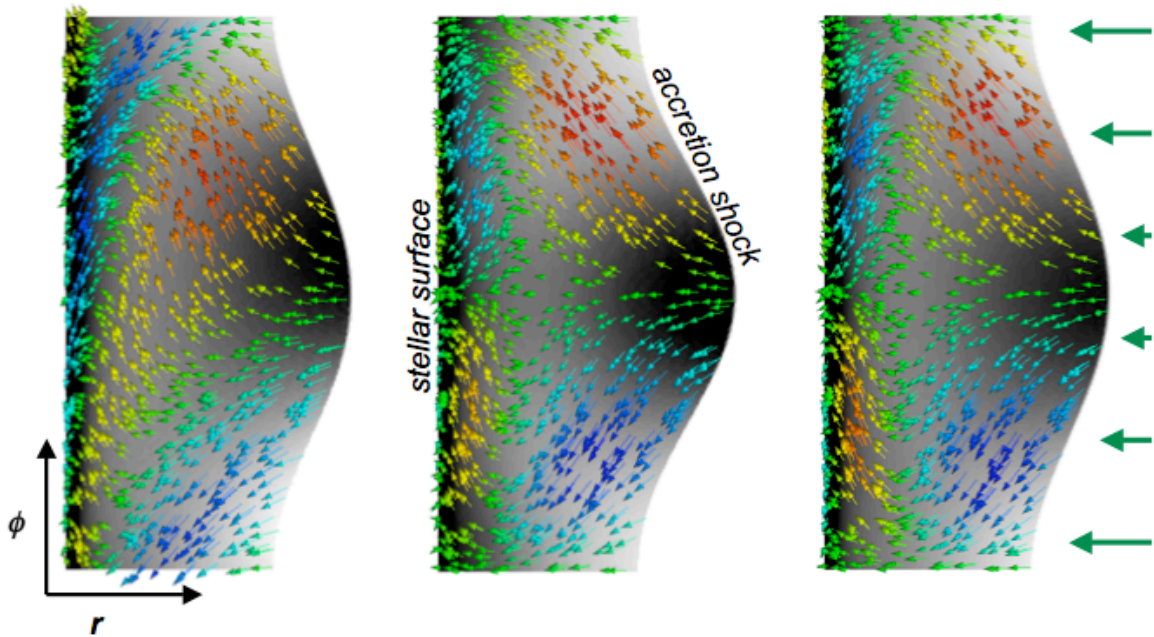

**Figure 3.** A comparison of the spiral (left) and sloshing (center) modes of the SASI. The last image is constructed by adding the spiral wave to a mirror image of itself, illustrating the relationship between the spiral and sloshing modes. These images are displayed in $r\phi$ coordinates, with the left edge corresponding to the surface of the accreting star and the right wavy edge corresponding to the accretion shock. The supersonic in-flow approaches the accretion shock horizontally from the right. The post-shock flow is represented by velocity vectors colored by the $\phi$-component of velocity. The background grey scale depicts the normalized pressure, $Pr^4$, and the flow vectors are colored by the angular velocity.

This spiral mode is consistent with the interpretation of the SASI as a global acoustic wave, but with the wave propagating continuously in the same direction around the circumference of the accretion shock rather than bouncing back and forth from pole to pole. The similarity of the sloshing and spiral modes is illustrated in Figure 3. In both cases the maximum pressure occurs at, or near, the most extended region of the accretion shock. In the case of the spiral mode, the high-pressure region is displaced slightly in the direction of propagation of the SASI wave (towards the top of the plot). In both modes the non-radial velocity is created by a combination of the shock obliquity and the local pressure gradient. These two effects combine to drive flow away from the most extended region of the shock.

Within the context of the two-dimensional geometry of these simulations, the sloshing mode can be thought of as a linear combination of two spiral waves moving in opposite directions. This is illustrated by the last image in Figure 3, where we have taken the data from a spiral mode and added it to a flipped version of itself (switch the direction of $\phi$ and changed the sign of the $\phi$-velocity). The result looks remarkably like the data from a simulation dominated by a sloshing mode. This decomposition does not directly carry over to the realistic case of a 3D spherical shock, but the close relationship of these two modes still exists in 3D. The sloshing mode is an acoustic wave that propagates along a symmetry axis from pole to pole, while the spiral mode is a similar acoustic wave that propagates continuously around an axis.

As in Paper II, we quantify the growth of these linear modes by tracking the amplitude of perturbed quantities such as the pressure or entropy. However, given the different grid geometry, here we compute the amplitude in different Fourier components instead of Legendre polynomials. The time evolution of the power in the $m=1$ mode for the model with $r_* = 0.3R_s$ is shown in Figure 4. For comparison we also plot the growth of the dominant axisymmetric mode ($l=1$) from a simulation where the SASI was excited by dropping an axisymmetric density perturbation onto the shock. The spiral mode grows with virtually the same rate as the sloshing mode. The power for the sloshing mode is modulated at the frequency of the standing wave, as the pressure is first high at one pole, then uniform, then high at the other pole. In contrast, the power for the spiral mode is (almost) monotonically increasing, as the pressure wave moves continuously around the circumference of the accretion cavity. The growth curve for the spiral mode does show some slight modulation at the rotation period of the wave. This is presumably the result of some small component of the sloshing mode excited along with the spiral mode (or alternatively, some small contribution of a spiral mode rotating in the opposite direction).

Our choice of $r_* = 0.3R_s$ is motivated by published supernova models, but in fact this ratio of the stalled shock radius to the radius of the PNS surface can vary significantly depending on the progenitor star and the details of the physics implemented in the supernova simulation. Moreover, this ratio may change throughout the evolution of the supernova as the PNS gradually shrinks and the accretion shock (perhaps?) expands. While our equilibrium models cannot address the effect of this evolution on the SASI, we can investigate the dependence of the SASI on the chosen fixed value of $r_*$. As seen in Figure 5, models with $r_* \sim 0.3R_s$ give the fastest growth, while models with a larger or smaller shock stand-off distance have smaller growth rates. This result is consistent with the axisymmetric results (Paper II) which show a peak in the growth rate of the $l=1$ mode near $r_* \sim 0.2R_s$.

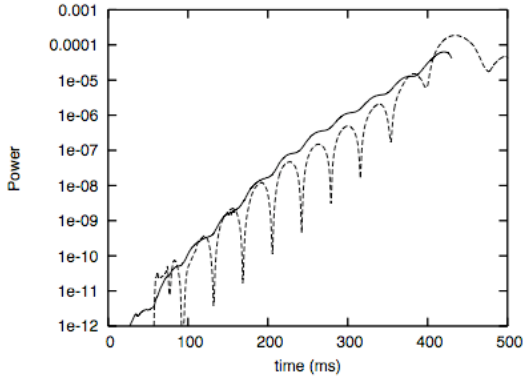
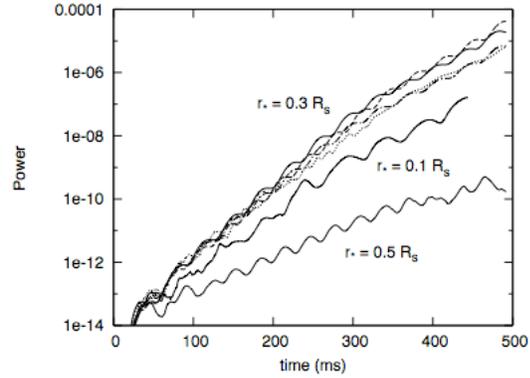

**Figure 4.** The growth of the SASI is quantified here by the power in the pressure perturbation for the $m=1$ spiral mode (solid line) and the $l=1$ sloshing mode (dashed line). The amplitude of the SASI is growing exponentially with a timescale of 40 ms.

**Figure 5.** The dependence of the SASI growth rate on the radius of the accreting PNS is illustrated with growth curves for models with different ratios of $r_*/R_s$. Also shown are several additional models with $r_* = 0.3R_s$ illustrating the dependence of these results on numerical resolution (dashed line has 225 radial zones, dot-dashed line has 900 zones) and inner boundary conditions (long dashed line [overlapping with the solid line] has zero tangential velocity at the boundary, dotted line has no mass taken off the grid at the boundary).

Multiple-armed spirals are also found to grow in our simulations. Figure 6 compares the growth curves for one-, two- and three-armed spirals. The growth rate is a monotonically decreasing function of $m$, consistent with the axisymmetric mode where the growth rate decreases with increasing $l$ (Paper II).

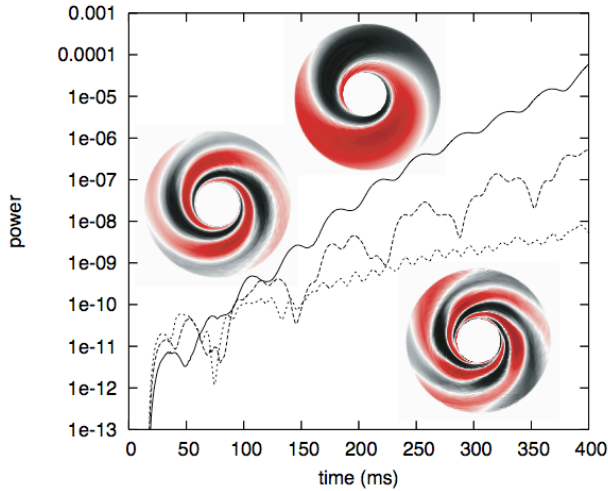

**Figure 6.** The growth rate of the SASI is a decreasing function of $m$, as shown by these results for $m = 1, 2$ and $3$.

Spiral waves are a familiar phenomenon in astrophysics, not only for forming the spiral arms of disk galaxies, but also arising in the context of accretion disks. In this latter case the spiral pattern is generated by radially propagating acoustic waves being sheared by the differential rotation of the Keplerian flow. In contrast, the SASI spiral pattern is generated by azimuthally propagating waves being sheared by the slower radial accretion flow. In both cases it is important to note that spiral waves carry angular momemtum (Larson 1990). In the context of accretion disks, nonlinear spiral waves can transport a significant flux of angular momentum radially outward through the disk (Blondin 2000). A similar situation arises with the spiral SASI waves, with positive (in the same sense as the SASI wave) angular momentum deposited in the outer region of the post-shock flow and negative angular momentum transported inwards. Because the net angular momentum in our simulations must remain zero (no external torques and no angular momentum in the radial flow at the outer boundary), the angular momentum in the spiral wave is matched by an equal and opposite angular momentum accreted onto the central star. As the amplitude of the spiral wave grows exponentially, so too does the angular momentum accreted onto the stellar surface. This exponential growth of the angular momentum of the accreting PNS is shown in Figure 7. Eventually the SASI becomes non-linear and other modes begin to affect the flow, resulting in a much more erratic accretion of angular momentum. Given that we are only simulating the flow in the equatorial plane, we have not attempted to estimate the total angular momentum deposited onto the PNS. We note, however, that similar 3D simulations suggest enough angular momentum can be deposited to leave behind a neutron star spinning with a period of 50 ms (Blondin and Mezzacappa 2006).

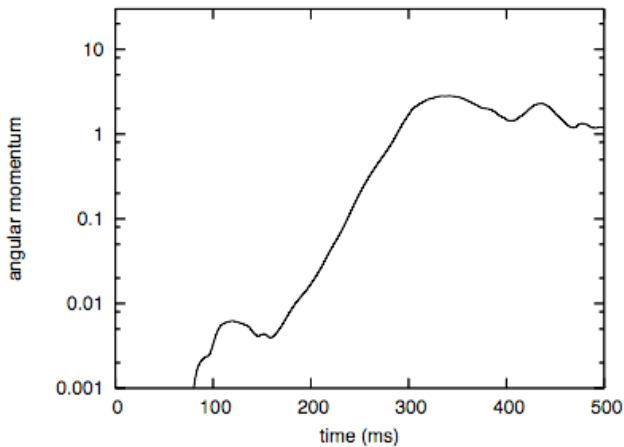

**Figure 7.** The net accumulated angular momentum (in arbitrary units) of the PNS for a simulation excited with the $m=1$ spiral mode. The SASI becomes nonlinear at a time of ~265 ms.

## 4. Non-Linear Evolution and Rotational Flow Driven by the SASI

The end of the epoch of linear growth of the spiral SASI modes is clearly delineated by the steepening of the interior wave into a shock. Because this wave is propagating along the accretion shock, its nonlinear form is referred to as a shock-shock (Whitham 1974). This shock-shock connects with the accretion shock at a shock triple-point, seen as a kink in the accretion shock in Figure 8. This shock triple point represents the leading edge of the non-linear SASI wave. In this 2D non-axisymmetric case, the $m=1$ mode has a single triple point that propagates in the same direction continuously around the surface of the accretion shock. In 3D this shock triple point is a line segment on the shock surface that extends roughly half way around the circumference of the shock (Blondin and Mezzacappa 2006).

A similar transition occurs for the axisymmetric sloshing mode of the SASI, but in these 2D simulations there are two shock triple-points (equadistant from the symmetry axis of the sloshing mode) that travel along the surface of the accretion shock from one pole to the other and back. In 3D these two points would correspond to a ring around the spherical shock. This ring would travel toward a pole where it would converge to a point, reflect off of itself, and propagate back around the surface of the shock to the other pole.

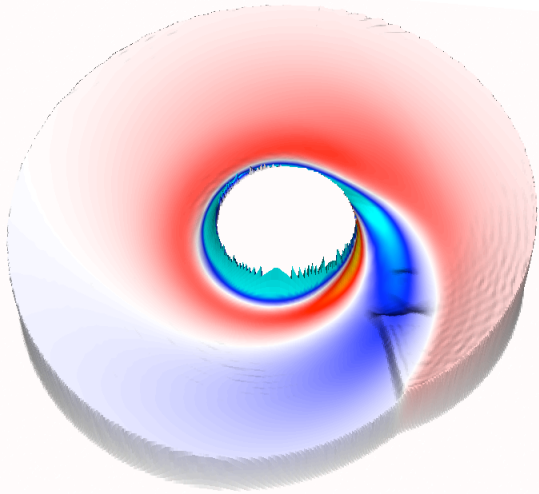 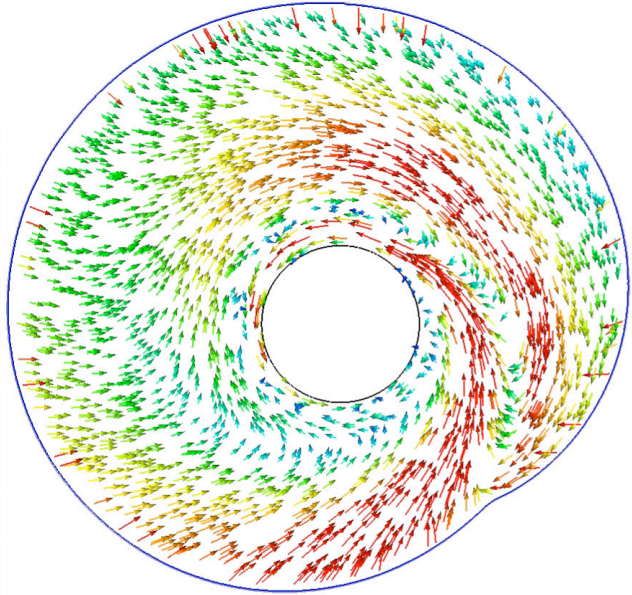

**Figure 8.** The structure of the spiral mode at a time of 277 msec, shortly after the end of the linear epoch. This image shows an elevated surface plot where the height is proportional to $Pr^4$ and the color represents the angular momentum density. The shock-shock is evident as a shallow ridge extending into the interior from the shock triple-point. Multiple weak reflected shocks can be seen along the spiral stream carrying angular momentum toward the surface of the accreting star.

**Figure 9.** The post shock flow pattern for the model shown in Figure 8 is illustrated with velocity vectors. The spiral wave is propagating in a clockwise direction, but a fast, narrow stream of gas stretches from the shock triple point to the surface of the accreting star, feeding the accretion of angular momentum with a counter-clockwise sense. The highly supersonic pre-shock flow is not included in order to improve visibility of the image.

The flow pattern generated by this shock structure is shown in Figure 9, corresponding to a time shortly after the transition to the non-linear regime. Here the SASI wave has 'crested', with the high point of the wave (the point of maximum shock radius as seen in Figure 3) racing forward (in a clockwise direction in this figure) and crashing on its own tail. The leading edge of this crashing wave is marked by the shock triple point, seen as a slight kink in the accretion shock in the bottom right of Figure 9. The upper-right hemisphere is the high-pressure part of the SASI wave where the accretion shock generates strong clockwise flow below the shock, moving with the SASI wave. Ahead of this wave (in a clockwise sense) the post-shock flow is moving in the opposite sense, converging into a narrow stream that feeds a strong counter-clockwise flow. This tail of the SASI wave is receding in radius, resulting in a lower relative shock velocity and lower post shock entropy. This stream of fast, low-entropy gas spirals down to the surface of the PNS, resulting in a large angular momentum accretion rate. These two counter-rotating streams illustrate the ability of the deformed accretion shock to separate the angular momentum of the accretion flow.

## 5. Discussion

The simulations presented in this paper have demonstrated that the non-axisymmetric modes of the SASI are, like their axisymmetric counterparts, linear instabilities that grow exponentially on a time scale of order 40 ms. The existence of these modes raises the concern that the dynamics of the stalled accretion shock in core collapse supernova will necessarily be incomplete in traditional 2D axisymmetric simulations. Moreover, 3D simulations of spherical accretion shocks are dominated by these spiral SASI modes at late time (Blondin 2005), so their forced absence in 2D simulations is a severe limitation. This word of caution is not necessarily restricted to the SASI, as other interesting dynamics may yet be discovered in 3D.

These spiral SASI modes are straight-forward to understand in the context of global acoustic waves. In this picture the spiral mode is an acoustic wave propagating in a purely azimuthal direction around the circumference of the post-shock cavity, immediately below the accretion shock. There is no evidence of any radial wave propagation anywhere in these simulations, contradicting the vortical-acoustic interpretation of the SASI (Foglizzo et al. 2006). In our 2D equatorial geometry the sloshing mode can be thought of as two oppositely moving spiral waves, just as a standing wave on a string can be thought of as two counter-propagating waves bouncing between fixed end-points. Eventually these interior waves steepen into shocks, forming a kink (a shock triple-point) in the otherwise smooth accretion shock. The strong non-radial flows in the interior are a direct result of the large deformation of the accretion shock in the vicinity of this triple point.

Perhaps the most intriguing consequence of these spiral SASI modes is the accretion of angular momentum onto the central PNS when the progenitor star is non-rotating. The distorted accretion shock generates counter-rotating flows with equal and opposite angular momentum. The flow moving in the opposite direction of the SASI wave is accreted onto the PNS, while the angular momentum associated with the SASI wave is presumably expelled in the subsequent supernova explosion. Using 3D simulations, Blondin and Mezzacappa (2006) found that enough angular momentum was accreted in this process to leave behind a neutron star with a spin period as fast as 50 ms. The spiral SASI provides the unexpected result that a non-rotating star with no external torques can leave behind a rapidly rotating neutron star.

This result of generating strong rotational flow in a non-rotating progenitor begs the question of how this mode is affected by rotation of the progenitor star. While we have not addressed this question in the present work, we note that initial 3D simulations that include progenitor rotation favor the development of the spiral mode over the sloshing mode seen in 2D simulations (Blondin and Mezzacappa 2006), and a linear perturbation analysis finds higher growth rates for non-axisymmetric modes in the presence of rotation (Laming 2006). We thus expect these results to carry over to the more general case of a rotating progenitor star. If the core is rotating rapidly, the dynamical instability of the self-gravitating, rotating core can also lead to an *m=1* spiral pattern, independent of the SASI (Ott et al. 2005). This coincidence raises the possibility that these two distinct physical mechanisms might feed off of each other, leading to an even stronger dominance of an *m=1* pattern than one would find in the case of just the SASI or the rotational instability alone.

The flow pattern generated by the spiral SASI modes is characterized by strong shear layers. Assuming there is at least some relic magnetic field present in the core material falling through the accretion shock, one can expect this shear to have a significant influence on the evolution of the field, and perhaps an equally important influence of the sheared field back on the flow. Akiyama et al. (2003) has argued that even weak magnetic fields can be dynamically important in the collapse of a rotating stellar core through the magneto-rotational instability. The spiral SASI can provide an alternative source of rotation and shear, extending the results of Akiyama et al. (2003) to include even slowly rotating progenitor stars.

The models presented here are designed to mimic the stalled accretion shock of the post-bounce phase in core collapse supernovae. Using these equilibrium models we have shown that the non-axisymmetric mode of the SASI can grow on a sufficiently short time scale to be relevant to the supernova mechanism, and that these modes can generate strong rotational flow in a spiral pattern, leading to the accretion of angular momentum onto the PNS. Given that the existence of the sloshing mode of the SASI has been confirmed in more extensive supernova models that include neutrino heating and the associated convection (Ohnishi et al. 2005), we expect our results here to carry over to comparable 3D supernova models that allow for non-axisymmetric modes.

This work was supported by the Terascale Supernova Initiative, funded by a SciDAC grant from the U.S. Department of Energy. We thank Anthony Mezzacappa for the many fruitful discussions over the course of this work, and John Hawley for suggesting the use of a 2D equatorial grid to study non-axisymmetric modes.

**REFERENCES**

Akiyama, S., Wheeler, J. C., Meier, D. L. & Lichtenstadt, I. 2003, ApJ, 584, 954
Blondin, J. M. 2000, New A, 5, 53
Blondin, J. M. 2005, in Journal of Physics Conf. Ser. 16, SciDAC 2005, ed. A. Mezzacappa (Philadelphia: IoP), 370
Blondin, J.M. & Mezzacappa, A. 2006, Nature, submitted
Blondin, J.M. & Mezzacappa, A. 2006, ApJ, 642, 401 (Paper II)
Blondin, J. M., Mezzacappa, A., & DeMarino, C. 2003, ApJ, 584, 971 (Paper I)
Buras, R., Rampp, M., Janka, H.-Th. & Kifonids, K. 2003, PhRvL, 90, 1101
Foglizzo, T., Galletti, P., Scheck, L. & Janka, H.-Th. 2006, ApJ, in press
Janka, H.-T., Buras, R., Kitaura Joyanes F.S., Marek, A. & Rampp, M. 2004, in Proc. 12th Workshop on Nuclear Astrophysics, ed. E. Müller & H. T. Janka (MPA/P14; Garching: MPI), 150
Laming, J. M. 2006, ApJ, submitted
Larson, R. B. 1990, MNRAS, 243, 588
Liebendoerfer, M. et al. 2001, PhRvD, 63, 3004
Mezzacappa, A. 2005, ARNPS, 55, 467-515
Ohnishi, N., Kotake, K. & Yamada, S. 2006, ApJ, 641,1018
Ott, C. D., Ou, S., Tohline, J. E. & Burrows, A. 2005, ApJL, 625, L119
Whitham, G. B. 1974, Linear and Nonlinear Waves (New York: Wiley)